\documentclass[amsmath,twocolumn,showpacs,showkeys]{revtex4}
\usepackage{amsmath}
\usepackage{amssymb}
\usepackage{latexsym}
\usepackage{amssymb}
\usepackage{epsfig}
\usepackage{natbib}
\usepackage{bm}
\usepackage{color}
\usepackage{rotating}
\def \ii{{\mathrm{i}}}

\def \d{{\mathrm{d}}}

\def \pd{\partial}

\def \e{{\mathrm{e}}}

\def\onedot{$\mathsurround0pt\ldotp$}
\def\cdddot#1{
  \mathbin{\vcenter{\baselineskip.67ex
    \hbox{\onedot}\hbox{\onedot}\hbox{\onedot}%
  }}%
}

\begin{document}
\title{Gradient modification of Newtonian gravity}
\author{Markus Lazar}
\email{lazar@fkp.tu-darmstadt.de}
\affiliation{Department of Physics,
        Darmstadt University of Technology,
        Hochschulstr. 6,    
        D-64289 Darmstadt, Germany}


\begin{abstract}
A second gradient generalization of Newtonian gravity is presented within the framework of 
gradient field theory. 
Weak nonlocality is introduced via first and second gradients of the gravitational field strength in the Lagrangian density. 
Gradient generalizations of the Poisson equation of Newtonian gravitation for the gravitational potential
and of the generalized Gauss law for the gravitational field strength  are presented.
Such a gradient modification of Newtonian gravity provides a straightforward regularization of Newtonian gravity removing the 
classical Newtonian singularities.
Finite gradient modifications of the gravitational potential energy and of the gravitational force law 
are constructed, with a possible connection to Yukawa interaction, and as suitable candidates for experimental tests
of Newton's inverse-square law at short distances.  
In addition, nonlocal gravity of exponential type is investigated and its relation to gradient gravity theory is given. 
\end{abstract}


\date{30 September 2020}

\maketitle

\section{Introduction}

During the last years, cosmological observations have shown that 70\% of all of the mass and energy of the Universe is related to a 
mysterious ``dark energy" leading to a repulsive gravitational effect (see also \citep{Kapner07,Hoyle}).
Only 5\% of the mass of the Universe is in the form of baryons, while 25\% is in the form of ``dark matter".  
The ``dark energy" and ``dark matter" play an important role in the evolution of the Universe.
New physics phenomena due to the ``dark energy", ``dark matter" and extra dimensions of M-theory could occur below the length scale associated with dark energy $\lambda_\text{d}\approx 85\, \mu$m 
and may modify the gravitational inverse-square law.

Moreover, the singularities in Newtonian gravity indicate the limits of applicability of the theory.
For this reason, they represent an important motivation to study generalized theories of gravitation 
(e.g. nonlocal gravity, gradient gravity, higher derivative quantum gravity) 
to make the theory free of singularities.
The characteristic singularities appearing for a point mass in classical Newtonian gravity are $1/R$- and $1/R^2$-singularities
in the Newtonian potential and Newtonian force, respectively.

Therefore, there is a strong interest for studying generalizations of and deviations from Newtonian gravity at small scales (see, e.g., \citep{Fischbach,Adelberger05,Murata}).
It should be emphasized that we are not concerned with the known deviations from Newtonian gravity in
strong gravitational fields or in relativistic systems at cosmic scales, which are described by
Einstein's general relativity and its generalizations. 
Our focus will be exclusively on possible deviations that arise in systems
that should be described by Newtonian physics at short distances.

From the theoretical point of view, 
there are interesting gradient and nonlocal modifications of gravity.
For instance, \citet{Treder} used a unified field theory of gravitation with long- and short-range interactions
which is related to the so-called fourth order gravity 
and found a non-singular gravitational potential of Bopp-Podolsky type (see also \citep{Treder2}). 
In particular, the short-range interaction gives rise to a repulsive force. 
\citet{HM} used a nonlocal modification of Newtonian gravity with a particular scalar nonlocal kernel in order to simulate dark matter (see also \citep{H10,Mashhoon}). 
On the other hand, 
higher-derivative gravity models are often used in quantum gravity (see, e.g., \citep{Stelle, Asorey,Modesto}). 

From the experimental point of view, 
until today, there are many investigations of Newton's inverse-square law at short distances (see, e.g., \citep{Hoyle04,Kapner07,Adelberger07,Murata,Adelberger20,Tan}). 
A standard parameterization of a violation of the  inverse-square law used in experimental tests
adds to the Newtonian gravitational potential a Yukawa term with strength $\alpha$ and range $\lambda$  for two masses $M$ and $M'$ 
(see, e.g., \citep{Fischbach,Hoyle04,Adelberger05}):
\begin{align}
\label{NY}
U=-\frac{GMM'}{R}\,\big[1+\alpha\, \e^{-R/\lambda}\big]\,,
\end{align}
where $G$ is the Newtonian gravitational constant.
In a test of the gravitational inverse-square law below the dark-energy length scale, \citet{Kapner07} found that any gravitational-strength ($|\alpha|=1$) Yukawa interaction must have $\lambda<56\, \mu$m.
A recent experimental test of \citet{Adelberger20} has found that any gravitational-strength Yukawa interaction must have $\lambda<38.6\, \mu$m.  
Therefore, based on the mentioned experimental data,  
characteristic length scale(s) of a modified Newtonian gravity should be in the order of $\mu$m or even smaller. 
The modified gravitational force based on Eq.~\eqref{NY} reads
\begin{align}
\label{FY}
{\bm F}=-\frac{GMM'}{R^2}\,\Big[1+\alpha\,\Big(1+\frac{R}{\lambda}\Big)\, \e^{-R/\lambda}\Big]\, \frac{\bm R}{R}\,.
\end{align}
It is noted that only for $\alpha=-1$, Eqs.~\eqref{NY} and \eqref{FY} are singularity-free.

Thus, the current experimental tests by \citet{Adelberger20} 
and \citet{Tan} of Newton's inverse-square law are in agreement with Newtonian gravity down to $52\, \mu$m
and $48\, \mu$m, respectively. 
However, Newton's inverse-square law of the gravitational force possesses a $1/R^2$-singularity which should be modified 
at short distances towards a singularity-free gravitational force expression. 
As always in theoretical physics, the range where the singularity becomes dominant shows that such theory is not valid at this range and must be modified or 
at least regularized.  
Therefore, the classical Newtonian singularities present in Newtonian gravity indicate the limits of the applicability of Newton's  theory of gravity at 
short distances. 
Non-singular versions of Newtonian gravity may be  nonlocal and gradient modifications of Newtonian gravity delivering easy-to-use singularity-free 
analytical expressions for the gravitational force (modified Newtonian force) and the modified Newtonian potential  depending on characteristic length scale parameters,
which might be used and determined in experiments for fitting data in the search and test of a modified Newtonian force or a modified Newtonian potential at short distances. 
The appearing characteristic length scale parameters determine the range of the modification in the near field, and from the mathematical point of view they have 
the meaning of regularization parameters.  
Gradient modification of Newtonian gravity is in full agreement with Newtonian gravity in the far field. 
In the near field, gradient modification of Newtonian gravity 
provides a straightforward regularization based on higher order partial differential equations. 
Hence, a gradient modification of Newtonian gravity provides a regularization of Newtonian gravity at short distances 
similar to the Pauli-Villars regularization in quantum electrodynamics.    
The regularized versions of the gravitational force and gravitational potential  given in this paper can be used and tested in short-range gravity experiments.

The aim of the present work is to derive a modified Newtonian gravity based on gradient field theory
which is a singularity-free generalized continuum theory valid at short distances. 
Such a gradient modification of Newtonian gravity is nothing but a straightforward regularized version of Newtonian gravity. 
In particular, using such a gradient gravity we find that gravity weakens at short distances, so that no singularity is created.
Moreover, we want to find the gradient modifications of the gravitational potential energy and of the gravitational force law, 
with a possible connection to a Yukawa interaction. 

The outline of this paper is as follows.
In Section~\ref{sec2}, the theory of 
second gradient modification of Newtonian gravity including the 
generalized Gauss law for gravity is presented. 
In Section~\ref{sec3}, we give the relevant 
Green function and its first gradient.
In Section~\ref{sec4}, the non-singular gravitational fields of 
a point mass, the modified gravitational potential energy
and modified gravitational force law
are computed in the framework of second gradient gravity.
The limits of those gravitational fields to the first gradient gravity  
and to the classical Newtonian gravity are given in Section~\ref{sec5}
and Section~\ref{sec6}, respectively. 
In Section~\ref{exp}, a nonlocal modification of exponential type of Newtonian gravity is investigated and the 
relation to gradient modifications of Newtonian gravity is given. 
The conclusions are given in Section~\ref{concl}.

\section{Second gradient modification of Newtonian gravity}
\label{sec2}
In this Section, we provide the field-theoretical framework 
of second gradient modification of Newtonian gravity.
The Lagrangian density for second gradient modification of Newtonian gravity is given by
\begin{align}
\label{L-grad2}
{\cal L}_{}&=
\frac{1}{8\pi G}\, 
\Big(\bm g \cdot \bm g 
+\ell_1^2 \nabla \bm g :\nabla \bm g 
+\ell_2^4 \nabla\nabla \bm g \mathbin{\vdots} \nabla\nabla \bm g\Big)
+\rho\Phi
\end{align}
with the notation:
$ \bm g \cdot\bm g =g_i g_i$,
$ \nabla \bm g :\nabla \bm g =\pd_j g_i \pd_j g_i$ and
$ \nabla\nabla \bm g \mathbin{\vdots} \nabla\nabla \bm g =\pd_k\pd_j g_i \pd_k \pd_j g_i$.
Here 
$\bm g $ denotes the gravitational field strength vector,
$\Phi$ is the gravitational potential, and 
$\rho$ is the mass density. 
Moreover, $\ell_1$ and $\ell_2$ are two (real) internal characteristic length scale parameters of second gradient modification of Newtonian gravity 
and $\nabla$ is the del operator.
In addition to the classical term, first and second spatial derivatives of the gravitational field strength $\bm g$ 
multiplied by the characteristic lengths $\ell_1$ and $\ell_2$ appear
in Eq.~\eqref{L-grad2} which describe a weak nonlocality in space.
In general, gradient gravity is a local theory with a finite number of derivatives of the gravitational field strength.

The gravitational field strength $\bm g$ 
can be written as gradient of a scalar potential, called the gravitational potential:
\begin{align} 
\label{g}
\bm g&=-\nabla \Phi\,.
\end{align}
Therefore, the gravitational field strength is irrotational and 
satisfies the gravitational Bianchi identity
\begin{align}
\label{BI}
\nabla\times \bm g&=0\,.
\end{align}

The Euler-Lagrange equation derived from the Lagrangian density~\eqref{L-grad2}
reads 
\begin{align}
\label{EL}
\frac{\delta {\cal L}}{\delta \Phi}\equiv
\frac{\pd {\cal L}}{\pd \Phi}
-\nabla \cdot \frac{\pd {\cal L}}{\pd(\nabla\Phi)}
&+ \nabla\nabla:\frac{\pd {\cal L}}{\pd(\nabla\nabla\Phi)}
\nonumber\\
&-\nabla\nabla\nabla\mathbin{\vdots}  \frac{\pd {\cal L}}{\pd(\nabla\nabla\nabla\Phi)}=0
\end{align}
and gives the inhomogeneous field equation of second gradient modification of Newtonian gravity
\begin{align}
\label{EL-1}
& L(\Delta)\,
\nabla\cdot \bm g= -4\pi G\,\rho\,,
\end{align}
where the differential operator of fourth order is given by
\begin{align}
\label{L-op}
 L(\Delta)=1-\ell_1^2 \Delta +\ell_2^4 \Delta^2\,.
\end{align}
Here $\Delta$ denotes the Laplacian.
Eq.~\eqref{EL-1} represents the generalized Gauss law for gravity.

By substituting Eq.~\eqref{g} into Eq.~\eqref{EL-1}, a modified Poisson equation follows 
for the gravitational potential $\Phi$:
\begin{align}
\label{phi-L}
 L(\Delta)\,\Delta\,\Phi&=4\pi G\, \rho\,,
\end{align}
which is a partial differential equation of sixth order. 

Using the factorization of differential operators of higher order into a product of differential operators of lower order, 
the differential operator of fourth order~\eqref{L-op}  
can be factorized into a product of two differential operators of second order,
namely two Helmholtz operators
with characteristic length scale parameters $a_1$ and $a_2$:
\begin{align}
\label{L-op-2}
L(\Delta)=\big(1-a_1^2\Delta\big)\big(1-a_2^2\Delta\big)
\end{align}
with
\begin{align}
\label{a1a2-1}
\ell_1^{2}&=a_1^{2}+a_2^{2}\, ,\\
\label{a1a2-2}
\ell_2^{4}&=a_1^{2}\, a_2^{2}\,
\end{align}
and 
\begin{align}
\label{a1-2}
a^{2}_{1,2}&=\frac{\ell_1^{2}}{2}\Bigg(1\pm\sqrt{1-4\,\frac{\ell_2^{4}}{\ell_1^{4}}}\Bigg)\,.
\end{align}
The differential operator~\eqref{L-op-2} may be called bi-Helmholtz operator.

The two length scales $a_1$ and $a_2$ might be real or complex.
The condition for the character, real or complex, of the two lengths 
$a_1$ and $a_2$ 
is the condition 
for the discriminant in Eq.~\eqref{a1-2}, $1-4\ell_2^4/\ell_1^4$, 
to be positive or negative.
Depending on the character of the two length scales $a_1$ and $a_2$,
it can be distinguished between three cases:
\begin{itemize}
\item[(1)] \,
$\ell_1^4>4\ell_2^4$\,: {\it real case}\\
The two length scales $a_1$ and $a_2$ are real and distinct and they read 
\begin{align}
\label{a1-2-2}
a_{1,2}&=\ell_1\,\sqrt{\frac{1}{2}\pm \frac{1}{2}\,\sqrt{1-4\left(\frac{\ell_2}{\ell_1}\right)^{\!4}}}\,,
\end{align}
satisfying the condition $a_1>a_2$.
The limit from second gradient theory to  first gradient theory is given by:  $\ell_2^4\rightarrow 0$. 
\item[(2)]\, 
$\ell_1^4=4\ell_2^4$\,: {\it real degenerate case}\\
The length scales $a_1$ and $a_2$ are real and equal
\begin{align}
 a_1 =a_2=\frac{\ell_1}{\sqrt{2}}=\ell_2\,.
 \end{align}  
There is no limit to first gradient theory.
\item[(3)]\, 
$\ell_1^4<4\ell_2^4$\,:  {\it complex conjugate case}
\\
The two length scales $a_1$ and $a_2$  are complex conjugate
\begin{align}
\label{c1-2-c}
a_{1,2}&=
A\pm\ii B\,,
\end{align}
 with
\begin{align}
\label{AB}
A=\ell_2\,\sqrt{\frac{1}{2}+\frac{\ell_1^2}{4\ell_2^2}}\,,\qquad
B=\ell_2\,\sqrt{\frac{1}{2}-\frac{\ell_1^2}{4\ell_2^2}}\,,
\end{align}
where $A>0$ and $B>0$. 
There is no limit to first gradient theory.
 \end{itemize}

\section{Green function in second gradient modification of Newtonian gravity}
\label{sec3}

In second gradient modification of Newtonian gravity, 
the necessary Green function $G^{L\Delta}$ of the sixth order differential operator 
$L(\Delta)\, \Delta$ is defined by 
\begin{align}
\label{PDE6}
 L(\Delta)\,
\Delta\, G^{L\Delta}(\bm R)=\delta(\bm R)\,,
\end{align}
where $\bm R= \bm r-\bm r'$ and $\delta$ is the Dirac delta-function. 
The Green function of Eq.~\eqref{PDE6} is given by (see, e.g., \citep{Lazar19,LL20})
\begin{align}
\label{GF}
G^{L\Delta}(\bm R)
&=-\frac{1}{4\pi R}\, f_0(R,a_1,a_2)\,,
\end{align}
where $f_0(R,a_1,a_2)$ is a characteristic auxiliary  function.

For the case~(1), 
the auxiliary  function reads 
\begin{align}
\label{f0}
f_0(R,a_1,a_2)=1-\frac{1}{a_1^2-a_2^2}\,
\Big[a_1^2 \e^{-R/a_1}-a_2^2 \e^{-R/a_2}\Big]\,.
\end{align}
The series expansion (near field) of the auxiliary function~\eqref{f0} reads as 
\begin{align}
\label{f0-ser}
f_0(R,a_1,a_2)&=\frac{1}{(a_1+a_2)}\, R+\mathcal{O}(R^3)\,.
\end{align}
Therefore, the function $f_0(R,a_1,a_2)$ 
regularizes up to a $1/R$-singularity towards a non-singular field expression. 
Indeed, the Green function~\eqref{GF} is non-singular and finite at $R=0$, namely 
\begin{align}
\label{G-BB-0}
G^{L\Delta}(0)=-\frac{1}{4\pi (a_1+a_2)}\,.
\end{align}

For the case~(2),  
the auxiliary function~\eqref{f0} becomes 
\begin{align}
\label{f0-2}
f_0(R,a_1,a_1)=1
-\bigg[1+\frac{R}{2a_1}\bigg]\, \e^{-R/a_1}
\end{align} 
with the near field behaviour
\begin{align}
\label{f0-ser-2}
f_0(R,a_1,a_1)&=\frac{1}{2a_1}\, R+\mathcal{O}(R^3)\,.
\end{align}

For the case~(3),  
the auxiliary function~\eqref{f0} reduces to
\begin{align}
\label{f0-3}
f_0(R,a_1,a_2)=1
- \bigg[\cos(b R)+\frac{A^2-B^2}{2AB}\,\sin(b R)\bigg] 
\, \e^{-a R}\,, 
\end{align}
which is a real quantity with $a=A/\ell_2^2$ and  $b=B/\ell_2^2$,
and the near field behaviour reads as 
\begin{align}
\label{f0-ser-3} 
f_0(R,a_1,a_2)&=\frac{1}{2A}\, R+\mathcal{O}(R^3)\,.
\end{align}

Therefore, the Green function~$G^{L\Delta}(R)$ is a real quantity and finite at $R=0$ for the three cases~(1), (2) and (3).

The first gradient of the Green function~\eqref{GF} reads
\begin{align}
\label{GF-grad}
\nabla G^{L\Delta}(\bm R)
&=\frac{1}{4\pi}\, \frac{\bm R}{R^3}\,
f_1(R,a_1,a_2)
\end{align}
with the auxiliary function for the case~(1)
\begin{align}
\label{f1}
f_1(R,a_1,a_2)=1
&-\frac{1}{a_1^2-a_2^2}\,
\Big[a_1^2 \e^{-R/a_1}-a_2^2 \e^{-R/a_2}\Big]
\nonumber\\
&-\frac{R}{a_1^2-a_2^2}\,
\Big[a_1 \e^{-R/a_1}-a_2 \e^{-R/a_2}\Big]\,.
\end{align}
The series expansion (near field) of the auxiliary function~\eqref{f1} reads as
\begin{align}
\label{f1-ser}
f_1(R,a_1,a_2)&=\frac{1}{3 a_1a_2 (a_1+a_2)}\, R^3+\mathcal{O}(R^4)\,.
\end{align}
One can see that the function $f_1(R,a_1,a_2)$ regularizes up to a
$1/R^3$-singularity towards a non-singular field expression. 

For the case~(2),  
the auxiliary function~\eqref{f1} becomes
\begin{align}
\label{f1-2}
f_1(R,a_1,a_1)=1-\bigg[1+\frac{R}{a_1}+\frac{R^2}{2a_1^2}\bigg]\, \e^{-R/a_1}\,
\end{align} 
with near field
\begin{align}
\label{f1-ser-2}
f_1(R,a_1,a_1)&=\frac{1}{6 a^3_1}\, R^3+\mathcal{O}(R^4)\,.
\end{align}

For the case~(3),  
the auxiliary function~\eqref{f1} reduces to 
\begin{align}
\label{f1-3}
f_1(R,a_1,a_2)=1
&- \bigg[\bigg(1+\frac{R}{2A}\bigg)\cos(b R)
\nonumber\\
&
+\bigg(\frac{A^2-B^2}{2AB}+\frac{R}{2B}\bigg)\sin(b R)\bigg]\, \e^{-a R}\,,
\end{align} 
which is a real quantity with the near field
\begin{align}
\label{f1-ser-3}
f_1(R,a_1,a_2)&=\frac{1}{6 A(A^2+B^2)}\, R^3+\mathcal{O}(R^4)\,.
\end{align}

Therefore, $\nabla G^{L\Delta}$ is real, non-singular and is zero at $R=0$ for the three cases~(1), (2) and (3).
A more detailed study of the three cases (1)--(3)  can be found in \citep{Lazar19} in the framework of gradient elasticity of bi-Helmholtz type.

The main feature of the case~(3) is the presence of oscillating terms. 
Depending on which quantity is greater in the pair $(a,b)$, the oscillating terms might be relevant in the Green function 
$G^{L\Delta}$ and its gradient.
The characteristic length of the Yukawa potential is $1/a$  
and the period of oscillating terms is $2\pi/b$. 
For $a>b$, the oscillations can be smooth and not dominant. 
For $a<b$, the period of oscillations is smaller than the range of the Yukawa factor
and an oscillation can give a relevant contribution; but this case changes $\ell_1^2$ to $-\ell_1^2$ in the Lagrangian density~\eqref{L-grad2}
and in the differential operator~\eqref{L-op}. 

It is interesting to mention, that a similar observation was done 
in the so-called sixth order gravity model with real and complex massive poles~\citep{Accioly} (see also \citep{MS,Gia}).

\section{Gravitational fields
in second gradient modification of Newtonian gravity}
\label{sec4}

In second gradient modification of Newtonian gravity,
the gravitational potential is the solution of the modified Poisson equation~\eqref{phi-L}
for a given mass density
\begin{align}
\label{Phi-g2}
\Phi&=4\pi G \, G^{L\Delta}*\rho\,,
\end{align}
where the Green function is given in Eq.~\eqref{GF}
and $*$ denotes the convolution in space.
The gravitational field strength vector is nothing but the (negative) gradient of Eq.~\eqref{Phi-g2} and it reads
\begin{align}
\label{g-g2}
\bm g&=-4\pi G \, \nabla G^{L\Delta}*\rho\,,
\end{align}
where the gradient of the Green function is given in Eq.~\eqref{GF-grad}.

\subsection{Gravitational fields of a point mass}

The  mass density of a point mass located at the position $\bm r'$ is given by
\begin{align}
\label{PM}
\rho=M\, \delta(\bm r-\bm r')\,,
\end{align}
where $M$ denotes the mass. 
$\bm r'$ is the position vector of the point mass and $\bm r$ is the field vector.

\begin{figure}[t]\unitlength1cm
\centerline{
\epsfig{figure=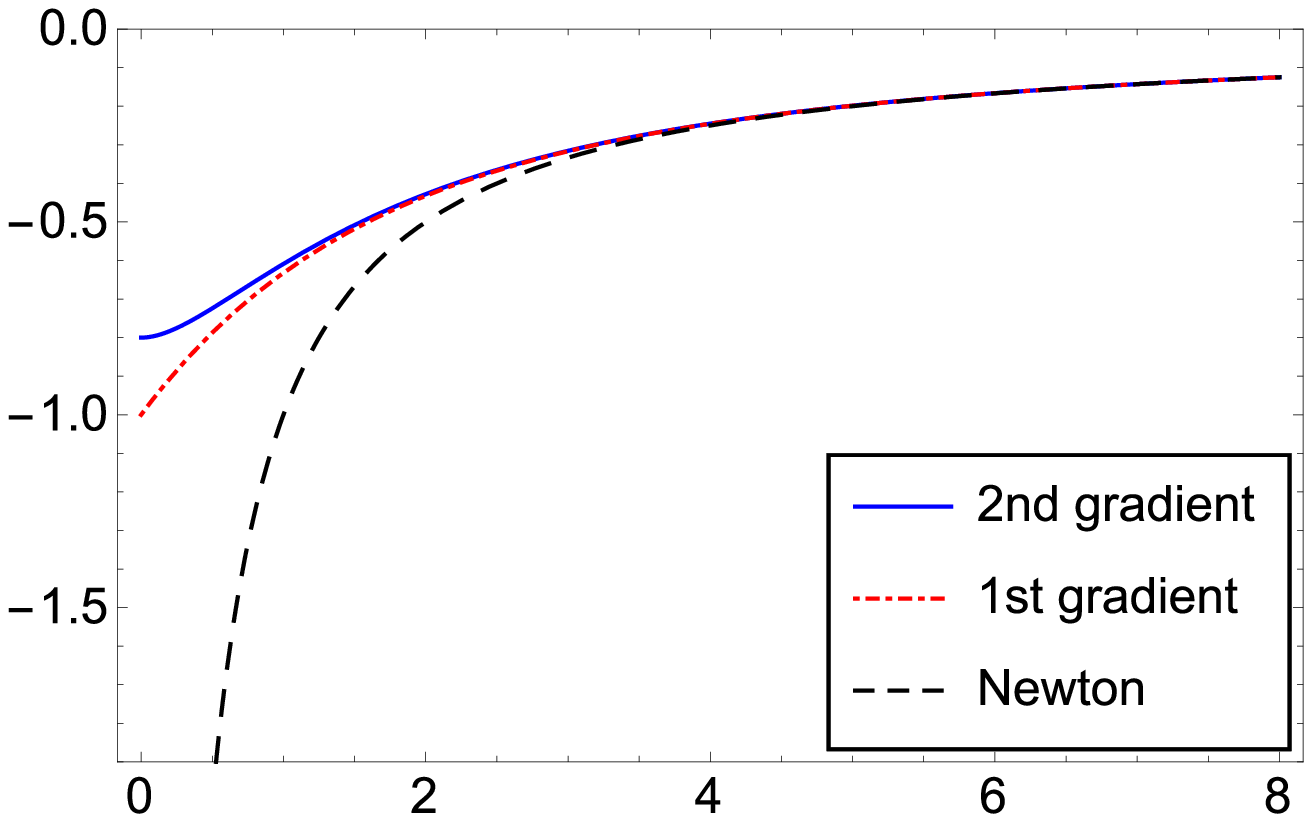,width=8cm}
\put(-4,0){$R/a_1$}
\begin{rotate}{90}
\begin{picture}(0,0)
\put(2.0,7.9){$\Phi a_1/GM$}
\end{picture}
\end{rotate}
}
\caption{Gravitational potential $\Phi$ of a point mass as a function of $R/a_1$
in second gradient Newtonian gravity (2nd gradient) for $a_1=4a_2$,
first gradient Newtonian gravity (1st gradient) 
and classical   Newtonian gravity (Newton).}
\label{fig:phi}
\end{figure}

On the one hand,
substituting Eq.~\eqref{PM} into Eq.~\eqref{Phi-g2} and performing the convolution, the gravitational potential of a point mass $M$ reduces to 
\begin{align}
\label{Phi-g3}
\Phi=4\pi G M \, G^{L\Delta}\,.
\end{align}
If we insert the Green function~\eqref{GF} into Eq.~\eqref{Phi-g3}, the explicit expression 
for  the modified gravitational potential of a point mass
reads in terms of the auxiliary function~\eqref{f0}
\begin{align}
\label{Phi-g4}
\Phi=-\frac{G M}{R}\, f_0(R,a_1,a_2)\,. 
\end{align}
Using the near field of $f_0$, Eq.~\eqref{f0-ser},  one can see that
the gravitational potential~\eqref{Phi-g4} is finite at $R=0$, namely  (see Fig.~\ref{fig:phi})
\begin{align} 
\label{Phi-g4-0}
\Phi(0)=-\frac{GM}{(a_1+a_2)}\,. 
\end{align}
In second gradient modification of Newtonian gravity, the gravitational potential~\eqref{Phi-g4} of a point mass 
is a superposition of a (long-range) Newtonian potential $-\frac{GM}{R}$ and
a (short-range) bi-Yukawa-type potential $\frac{GM}{(a_1^2-a_2^2)R}  [a_1^2 \e^{-R/a_1}-a_2^2 \e^{-R/a_2}]$
leading to a non-singular short-distance modification of the classical Newtonian potential of a mass $M$ 
(see also Fig.~\ref{fig:phi}). 
The cancellation or regularization of the $1/R$-Newtonian singularity in the gravitational potential is due to the opposite signs of the Newtonian potential and
the bi-Yukawa-type potential.
Therefore, Eq.~\eqref{Phi-g4} is the singularity-free modified Newtonian potential
in second gradient modification of Newtonian gravity.

Moreover, it is interesting to note that the modified Newtonian potential~\eqref{Phi-g4} with two free parameters
is a particular version of the  modified Newtonian potential  with four free parameters
obtained by~\citet{Accioly} in sixth order gravity.

On the other hand,
the gravitational field strength~\eqref{g-g2} of a point mass~\eqref{PM} reduces to 
\begin{align}
\label{g-g3}
\bm g=-4\pi G M\, \nabla G^{L\Delta}\,.
\end{align}
Using Eq.~\eqref{GF-grad}, Eq.~\eqref{g-g3} becomes
\begin{align}
\label{g-g4}
\bm g=-G M\, \frac{\bm R}{R^3}\, f_1(R,a_1,a_2)\,.
\end{align}
Using the near field of $f_1$, Eq.~\eqref{f1-ser},  one can see that
the gravitational field strength~\eqref{g-g4} is zero at $R=0$.
Due to its spherical symmetry, the gravitational field strength~\eqref{g-g4} of a point mass reduces to
\begin{align}
\label{g-g5}
\bm g=-\frac{G M}{R^2}\, f_1(R,a_1,a_2)\, \bm e_R\,,
\end{align}
where $\bm e_R=\bm R/R$ is the radial unit vector. 
In general, the  gravitational field strength~\eqref{g-g5} of a point mass is  non-singular and possesses 
an extremum value near the origin.

\subsection{Modified gravitational potential energy}

The gravitational field energy (potential energy) is given by
\begin{align}
\label{Ug}
U_{\text{g}}&=-\frac{1}{4\pi G} \int\limits_{\Bbb R^3}  
\Big(\bm g \cdot \bm g +\ell_1^2 \nabla \bm g :\nabla \bm g 
+\ell_2^4 \nabla\nabla \bm g \mathbin{\vdots} \nabla\nabla \bm g 
\Big) \text{d} V\nonumber\\
&=-\frac{1}{4\pi G} \int\limits_{\Bbb R^3}   L(\Delta) \bm g \cdot  \bm g\,  \text{d} V\nonumber\\
&=-\frac{1}{4\pi G} \int\limits_{\Bbb R^3}   L(\Delta) \nabla \cdot \bm g \,  \Phi\,  \text{d} V\nonumber\\
&=\int\limits_{\Bbb R^3}   \rho \,\Phi\,  \text{d} V\,,
\end{align}
where we have used integration by parts, Eq.~\eqref{EL-1}, and that the surface terms vanish at infinity.
Substituting Eq.~\eqref{PM} into Eq.~\eqref{Ug},
the  modified Newtonian gravitational interaction energy  of the two point masses $M$ and $M'$ becomes
\begin{align}
\label{UMM}
U_{MM'}&=M \Phi_{M'}\nonumber\\
&=-\frac{GMM'}{R}\, f_0(R,a_1,a_2)\,,
\end{align}
which is finite due to the non-singular potential~\eqref{Phi-g4} (see also Fig.~\ref{fig:phi}).

\subsection{Modified gravitational force law}
We proceed to the determination of the short-distance behavior of the modified force law.

\begin{figure}[t]\unitlength1cm
\centerline{
\epsfig{figure=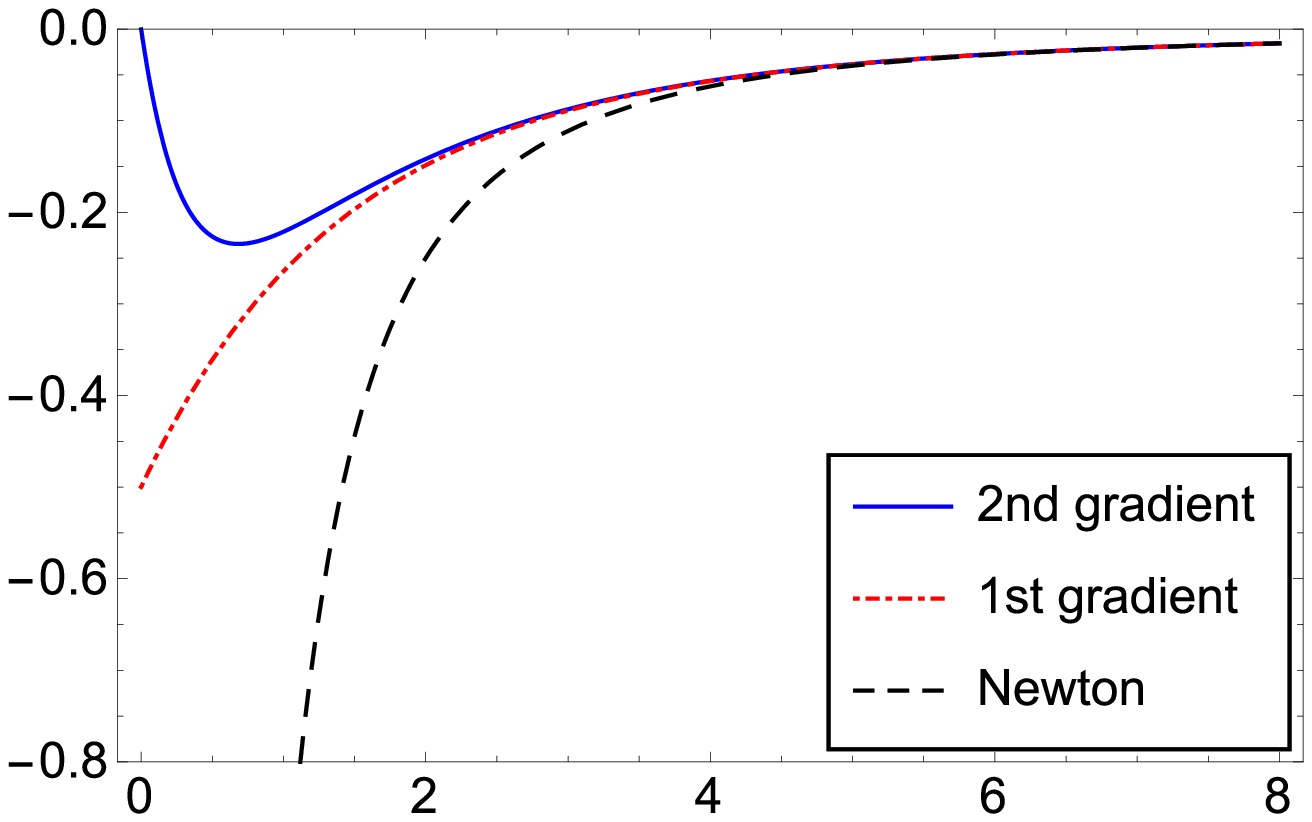,width=8cm}
\put(-4,0){$R/a_1$}
\begin{rotate}{90}
\begin{picture}(0,0)
\put(2.0,7.9){$F a_1^2/GMM'$}
\end{picture}
\end{rotate}
}
\caption{Gravitational force $F$ between two point masses as a function of $R/a_1$
in second gradient Newtonian gravity (2nd gradient) for $a_1=4a_2$,
first gradient Newtonian gravity (1st gradient) 
and classical   Newtonian gravity (Newton).}
\label{fig:F}
\end{figure}

In second gradient modification of Newtonian gravity, after a straightforward calculation the gravitational force reads as
\begin{align}
\label{F-N}
\bm{{F}}=\int\limits_{\Bbb R^3} \rho\, \bm g \, \d V\,,
\end{align}
which is the force acting on a mass distribution of density $\rho$ in presence of a gravitational field strength $\bm g$. 
Substituting Eqs.~\eqref{PM} and \eqref{g-g4} into Eq.~\eqref{F-N},
the gradient modification of Newton's inverse-square law of gravity is found as
\begin{align}
\label{FMM}
{\bm{F}}_{MM'}&=M \bm g_{M'}\nonumber\\
&=-G MM'\, \frac{\bm R}{R^3}\, f_1(R,a_1,a_2)\,.
\end{align}
The force~\eqref{FMM} is zero at $R=0$ and non-singular. 
Using  the radial unit vector $\bm e_R=\bm R/R$, the gradient modification of Newton's inverse-square law of gravity
reduces to (${\bm F}_{MM'}=F\, \bm e_R$)
\begin{align}
\label{FMM2}
{\bm F}_{MM'}=-\frac{G MM'}{R^2}\, f_1(R,a_1,a_2)\, \bm e_R\,.
\end{align}
The force~\eqref{FMM2} is plotted in Fig.~\ref{fig:F}.
Eq.~\eqref{FMM2} represents the attractive central conservative force 
acting on the point mass $M$ at $\bm r$ due to the presence of the point mass $M'$ at $\bm r'$.
It is interesting to note that the force~\eqref{FMM2} is a linear superposition of the attractive (long-range)
Newtonian force $-\frac{G MM'}{R^2}$ and a repulsive (short-range) 
bi-Yukawa-type force with spatial decay lengths $a_1$ and $a_2$. 
The cancellation or regularization of the $1/R^2$-Newtonian singularity in the gravitational force is due to the opposite signs of the classical 
Newtonian force and the bi-Yukawa-type force.
The exponential decay in the bi-Yukawa term originates from
the fading of spatial memory or weak nonlocality significant at short distances.
At small scales ($R<4a_1$), Newton's inverse-square law of gravity is strongly modified 
due to the bi-Yukawa-type force. 
Thus, objects separated by less than this distance ($R<4a_1$) would feel a reduced gravitational attraction,
and the force vanishes at $R=0$ (see Fig.~\ref{fig:F}). 
May this repulsive force be related with the mysterious ``dark energy"?
On the other hand, on the scales where $R>4a_1$ the bi-Yukawa-type force can be neglected and the force of gravity
is then essentially Newtonian. 
The modified gravitational force law~\eqref{FMM2} might be used for the investigation of a violation of Newton's inverse-square law at short distances. 
When a test particle is approaching $R=0$, the gravitational force~\eqref{FMM2} applied to it tends to zero 
because the repulsive bi-Yukawa force chancels the attractive Newtonian force (see Fig.~\ref{fig:F}).

Last but not least, it is interesting to note that the short-distance modification of Newton's force law based on Eq.~\eqref{FMM2} (see Fig.~\ref{fig:F})
possesses a similar form as the modified force law in Sundrum's ``fat graviton" model \citep{Sundrum}.

\section{Gravitational fields in first gradient modification of Newtonian gravity}
\label{sec5}

Now, we perform the limit from second gradient modification of Newtonian gravity
to first gradient modification of Newtonian gravity
which is given by: $\ell^4_2\rightarrow 0$, and therefore $a_1\rightarrow\ell_1$, 
$a_2\rightarrow0$.
In this limit, the auxiliary functions~\eqref{f0} and (\ref{f1}) reduce to
\begin{align}
\label{f0-H}
f_0(R,\ell_1)&=1-\e^{-R/\ell_1}\,,\\
\label{f1-H}
f_1(R,\ell_1)&=1-\bigg[1+\frac{R}{\ell_1}\bigg]\,\e^{-R/\ell_1}\,.
\end{align} 
The near fields of the auxiliary
functions~\eqref{f0-H} and \eqref{f1-H} read 
\begin{align}
\label{f0-H-ser}
f_0(R,\ell_1)&=\frac{1}{\ell_1}\, R-\frac{1}{2 \ell_1^2}\, R^2
+\mathcal{O}(R^3)\,,\\
\label{f1-H-ser}
f_1(R,\ell_1)&=\frac{1}{2\ell_1^2}\, R^2-\frac{1}{3 \ell_1^3}\, R^3+\mathcal{O}(R^4)\,.
\end{align}
In Eqs.~\eqref{f0-H-ser} and \eqref{f1-H-ser},
it can be seen that the function $f_0(R,\ell_1)$ regularizes up to a
$1/R$-singularity towards a  non-singular field expression, and
the function $f_1(R,\ell_1)$ 
regularizes up to a $1/R^2$-singularity towards a non-singular field expression.

In this limit, the modified gravitational potential~\eqref{Phi-g4} reduces to
\begin{align}
\label{Phi-H1}
\Phi=-\frac{GM}{R}\,  f_0(R,\ell_1)\,,
\end{align}
which is finite at $R=0$, namely  (see Fig.~\ref{fig:phi})
\begin{align} 
\label{Phi-H2}
\Phi(0)=-\frac{GM}{\ell_1}\,. 
\end{align}
In first gradient modification of Newtonian gravity, the gravitational potential~\eqref{Phi-H1} of a point mass 
is a superposition of the (long-range) Newtonian term $-\frac{GM}{R}$ and
a (short-range) Yukawa term $\frac{GM}{R} \e^{-R/\ell_1}$
leading to a non-singular short-distance modification of the classical Newtonian potential
(see Fig.~\ref{fig:phi}).
Moreover, it is noted that Eq.~\eqref{Phi-H1} is in agreement with the expression
given by~\citet{Treder} obtained in his unified field theory of gravitation gravitation with long- and short-range interactions,
which is a particular version of the so-called fourth order gravity. 

Note that  the modified Newtonian potential of Bopp-Podolsky type~\eqref{Phi-H1}  with one free parameter
is a particular version of the finite ``Stelle"-potential which is the modified Newtonian potential with two free parameters in fourth order gravity 
(see, e.g., \citep{Stelle,Stelle2,Treder2}).

Moreover, the gravitational field strength vector~\eqref{g-g5}
simplifies to
\begin{align}
\label{g-H1}
\bm g=- \frac{GM}{R^2}\, f_1(R,\ell_1)\, \bm e_R\,.
\end{align}
Using Eq.~\eqref{f1-H-ser},
one can see that the  gravitational field strength vector of a point mass is finite 
at the origin, namely 
\begin{align}
\label{g-H1-0}
\bm g(\bm R)=-\frac{GM}{2\ell_1^2}\, \bm e_R\, + \mathcal{O}(R).
\end{align}

The modified gravitational potential energy~\eqref{UMM} for  two point masses $M$ and $M'$ reduces to
\begin{align}
\label{UMM3}
U_{MM'}=-\frac{GMM'}{R}\, f_0(R,\ell_1)\,.
\end{align}
The modified gravitational potential~\eqref{UMM3} is a linear superposition of the (long-range)
Newtonian potential term $-\frac{G MM'}{R}$ and a  (short-range) 
Yukawa potential term $\frac{GMM'}{R} \, \e^{-R/\ell_1}$
with spatial decay length $\ell_1$. 
It is interesting to note that if we compare~\eqref{UMM3} with \eqref{NY}, we find 
that the  modified gravitational potential~\eqref{UMM3} is the only singularity-free version of Eq.~\eqref{NY}
with $\alpha=-1$ and $\lambda=\ell_1$. 

The modified force law~\eqref{FMM2} becomes
\begin{align}
\label{FMM3}
{\bm F}_{MM'}=-\frac{G MM'}{R^2}\, f_1(R,\ell_1)\, \bm e_R\,.
\end{align}
The force~\eqref{FMM3} is plotted in Fig.~\ref{fig:F}. 
Note that the force~\eqref{FMM3} is finite at $R=0$.
The modified force~\eqref{FMM3} is a linear superposition of the attractive (long-range)
Newtonian force $-\frac{G MM'}{R^2}$ and a repulsive (short-range) 
Yukawa-type force $\frac{GMM'}{R^2} \big[1+\frac{R}{\ell_1}\big] \e^{-R/\ell_1}$.
In the near field, the modified force~\eqref{FMM3} is stronger than the modified force~\eqref{FMM2} which becomes zero at $R=0$. 
Note that the  modified force law~\eqref{FMM3} is the singularity-free version of Eq.~\eqref{FY}
with $\alpha=-1$ and $\lambda=\ell_1$. 
Therefore, in first gradient modification of Newtonian gravity,
the strength of the Yukawa term in Eqs.~\eqref{NY} and \eqref{FY} is fixed to be $\alpha=-1$
to have singularity-free gravitational fields.

\section{Gravitational fields in classical Newtonian gravity}
\label{sec6}

The limit from first gradient modification of Newtonian gravity
to classical Newtonian gravity is given by $\ell_1\rightarrow 0$.

From Eqs.~\eqref{Phi-H1} and \eqref{g-H1}, 
the classical gravitational potential and classical gravitational field strength
of a point mass are recovered
\begin{align}
\label{Phi-N}
\Phi=-\frac{GM}{R}
\end{align}
and
\begin{align}
\label{g-N}
\bm g=- \frac{GM}{R^2}\,  \bm e_R\,.
\end{align}

\section{Nonlocal gravity of exponential type}
\label{exp}
In this Section, we consider a nonlocal gravity of exponential type and its relation to gradient modification of gravity. 
In particular, the nonlocal modification of Newtonian gravity is considered. 

For a nonlocal modification of  Newtonian gravity, the Lagrangian density reads as
\begin{align}
\label{L-nonl}
{\cal L}_{}&=
\frac{1}{8\pi G}
\int\limits_{\Bbb R^3}
K(\bm x-\bm y)\,
 \bm g(\bm x) \cdot 
 \bm g(\bm y)\, 
\d \bm y +\rho\Phi\,,
\end{align}
where $K(\bm x-\bm y)$ denotes the so-called nonlocal kernel function or form factor.
Following Efimov (e.g.~\citep{Efimov}), we consider in the static case a form factor obtained by acting an entire function of the Laplacian  
on the Dirac delta-function
\begin{align}
\label{kernel}
K(\bm x-\bm y)=L(\Delta) \, \delta (\bm x-\bm y)\,.
\end{align}
Using the form factor~\eqref{kernel}, the Lagrangian density~\eqref{L-nonl} reduces to
\begin{align}
\label{L-nonl-2}
{\cal L}_{}&=
\frac{1}{8\pi G}\,
 \bm g\cdot    (L(\Delta)\,\bm g) +\rho\Phi\,. 
\end{align}
In nonlocal gravity of exponential type, an exponential operator
is used (see, e.g., \citep{Biswas,Modesto,Modesto15})
\begin{align}
\label{L-exp}
 L(\Delta)=\e^{-\ell^2\Delta}\,,
\end{align}
where $\ell$ is the characteristic length scale of the nonlocal theory. 
This differential operator is of infinite order, characteristic for a nonlocal theory. 
Eventually, the exponential operator may be represented  as an infinite series in power of $\Delta$:  
\begin{align}
\label{L-exp2}
 L(\Delta)=\sum_{n=0}^\infty \frac{(-1)^n}{n!}\, \ell^{2n}\Delta^n\,.
\end{align}

\begin{figure}[t]\unitlength1cm
\centerline{
\epsfig{figure=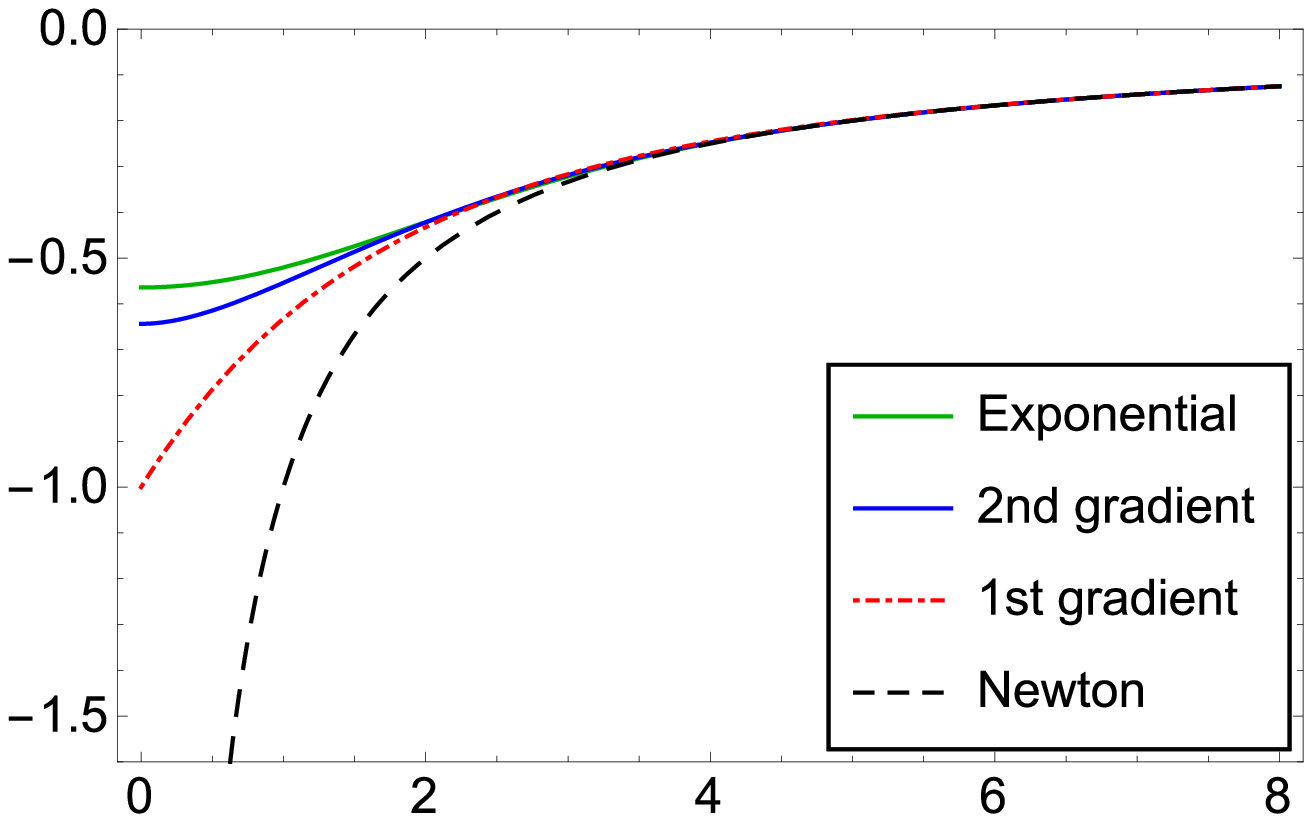,width=8cm}
\put(-4.0,0){$R/\ell$}
\begin{rotate}{90}
\begin{picture}(0,0)
\put(2.0,7.9){$\Phi \ell/GM$}
\end{picture}
\end{rotate}
}
\caption{Gravitational potential $\Phi$ of a point mass as a function of $R/\ell$
in nonlocal gravity of exponential type (Exponential),
second gradient Newtonian gravity (2nd gradient) for $\ell_1^4=2 \ell_2^4$,
first gradient Newtonian gravity (1st gradient) 
and classical   Newtonian gravity (Newton).}
\label{fig:phi-exp}
\end{figure}

\begin{figure}[t]\unitlength1cm
\centerline{
\epsfig{figure=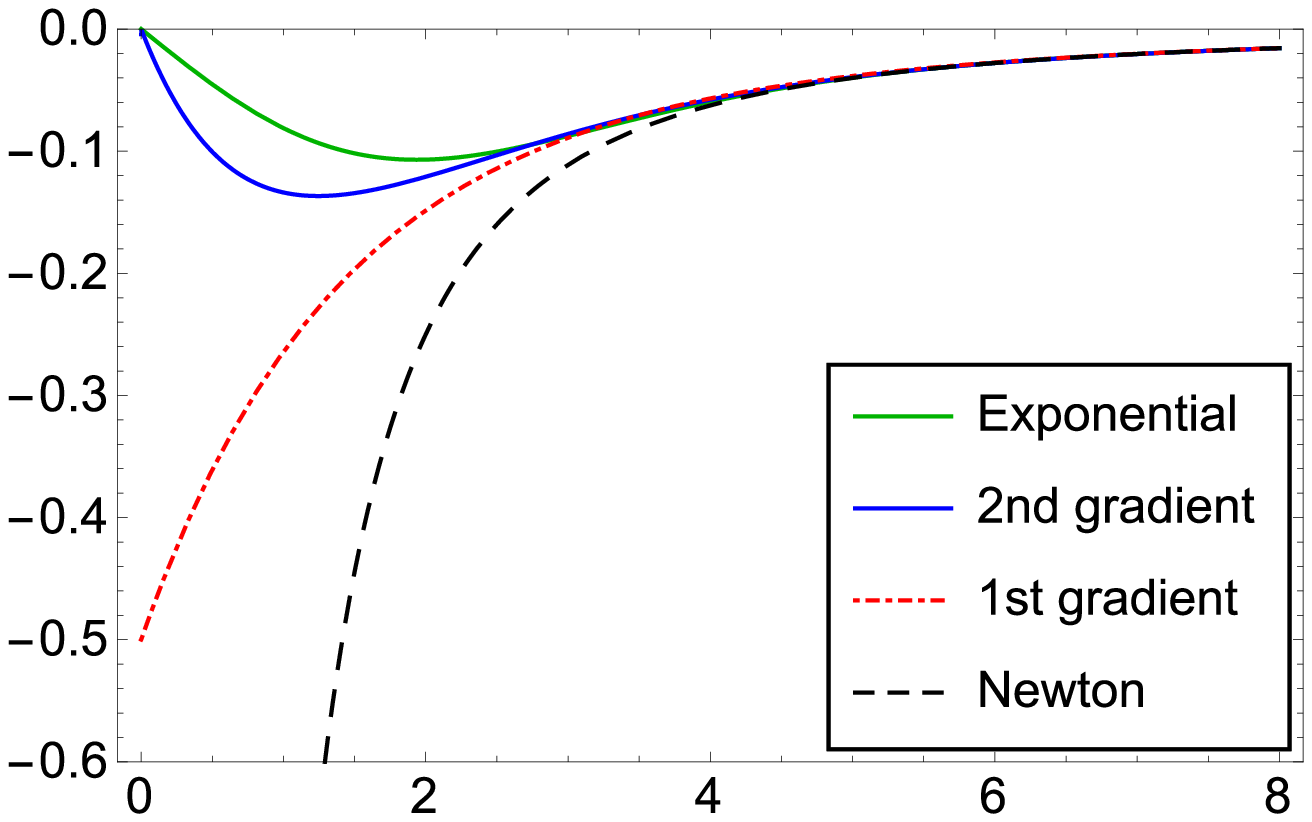,width=8cm}
\put(-4.0,0){$R/\ell$}
\begin{rotate}{90}
\begin{picture}(0,0)
\put(2.0,7.9){$F \ell^2/GMM'$}
\end{picture}
\end{rotate}
}
\caption{Gravitational force $F$ between two point masses as a function of $R/\ell$
in nonlocal gravity of exponential type (Exponential), 
second gradient Newtonian gravity (2nd gradient) for $\ell_1^4=2 \ell_2^4$,
first gradient Newtonian gravity (1st gradient) 
and classical   Newtonian gravity (Newton).}
\label{fig:F-exp}
\end{figure}

For the  exponential differential operator~\eqref{L-exp}, the solution of Eq.~\eqref{phi-L}  reads for the modified Newton potential \citep{Biswas,Modesto,Modesto15}
\begin{align}
\label{Phi-exp}
\Phi=-\frac{GM}{R}\,  \text{erf}\left(\frac{R}{2\ell}\right)\,,
\end{align}
where $\text{erf}$ denotes the  error function (also called the Gauss error function).
The modified Newton potential~\eqref{Phi-exp} is finite at $R=0$, namely  (see Fig.~\ref{fig:phi-exp})
\begin{align} 
\label{Phi-exp2}
\Phi(0)=-\frac{GM}{\sqrt{\pi}\ell}\,. 
\end{align}
The corresponding modified gravitational field strength reads as 
\begin{align}
\label{g-exp}
\bm g=- \frac{GM}{R^2} \left( \text{erf}\left(\frac{R}{2\ell}\right)-\frac{R}{\sqrt{\pi}\ell}\, \e^{-R^2/(4\ell^2)}\right) \bm e_R\,,
\end{align}
which is non-singular, and in particular it is zero at $R=0$. 
Moreover, the modified force law of nonlocal gravity of exponential type  becomes
\begin{align}
\label{FMM-exp}
{\bm F}_{MM'}=- \frac{GMM'}{R^2} \left( \text{erf}\left(\frac{R}{2\ell}\right)-\frac{R}{\sqrt{\pi}\ell}\, \e^{-R^2/(4\ell^2)}\right) \bm e_R\,,
\end{align}
which is plotted in Fig. \ref{fig:F-exp}.

The series expansion of the exponential operator~\eqref{L-exp} up to the order $n=2$  in Eq.~\eqref{L-exp2} 
is given by 
\begin{align}
\label{L-exp3}
 L(\Delta)=1-\ell^2 \Delta +\frac{1}{2}\, \ell^4 \Delta^2
 +\mathcal{O}(\Delta^3)\,.
\end{align}
In Eq. \eqref{L-exp3}, it can be seen
that the differential operator~\eqref{L-op} in second gradient modification of Newtonian gravity can be considered as truncated Taylor expansion 
of the exponential operator~\eqref{L-exp}.
If we compare Eq. \eqref{L-op} and Eq. \eqref{L-exp3}, we find the relation for the
length scales
\begin{align}
\label{rel-L}
\ell_1^2=\ell^2\,,\qquad
\ell_2^4=\frac{1}{2}\,\ell^4
\end{align}
and therefore, it yields 
\begin{align}
\label{rel-l}
\ell_1^4=2 \ell_2^4\,,
\end{align}
which belongs to the case (3) with complex conjugate length scales of the two Helmholtz operators in second gradient modification of Newtonian gravity.

Therefore, 
classical Newton gravity is the expansion of order $n=0$ of nonlocal gravity of exponential type.
First gradient modification of Newtonian gravity is the expansion of order $n=1$ of nonlocal gravity of exponential type.
Furthermore, second gradient modification of Newtonian gravity (case (3) with the relation for the two length scales \eqref{rel-l})
is the expansion of order $n=2$ of nonlocal gravity of exponential type. 
In this manner, second gradient modification of Newtonian gravity with complex conjugate length scales satisfying the condition~\eqref{rel-l}
can be understood as a proper approximation of nonlocal gravity of exponential type 
(see Figs. \ref{fig:phi-exp} and \ref{fig:F-exp}).

\section{Conclusions}
\label{concl}

Starting from first principles of field theory, 
a gradient modification of second order of Newton's theory of gravitation has been derived. 
In such a gradient theory, the gravitational potential field is local, but satisfies a partial differential equation of sixth order 
including internal characteristic length scales determining the range of modification of Newtonian gravity.
Moreover, the modified gravitational fields obtained in gradient theory are singularity-free and possess a modification 
at short distance and in the far field they converge to the Newtonian gravitational fields.   
The cancellation of the Newtonian singularities is due to the opposite signs of a Newtonian term and
a bi-Yukawa term.
The analytical expressions for the modified gravitational potential energy and the modified gravitational force law
obtained in the framework of gradient field theory can be used for testing Newton's inverse-square law at short distances
and for searching effects of non-Newtonian gravity in the order of micrometers or even smaller. 
The modified gravitational potentials obtained in second gradient modification and in first gradient modification of Newtonian gravity 
correspond to the modified Newtonian potentials of particular versions of sixth order and fourth order gravity, respectively. 
Therefore,  second gradient modification of Newtonian gravity
is the Newtonian limit of a particular version  
of sixth order gravity representing a simple version of quantum gravity which is local, superrenormalizable
and, in the case of complex massive poles, can be unitary. 

In addition, nonlocal modification of exponential type of Newtonian gravity and the corresponding gravitational fields, which are singularity-free,
have been studied. The connection between such nonlocal gravity of exponential type and gradient modification of 
Newtonian gravity has been given. 
It turned out that second gradient modification of Newtonian gravity with complex conjugate length scales
represents a very good approximation of nonlocal gravity of exponential type. 

Due to the comparison of second gradient modification of Newtonian gravity with sixth order gravity
and nonlocal gravity of exponential type, it can be concluded that 
second gradient modification of Newtonian gravity with complex conjugate length scales 
seems to be the most interesting case for a gradient modification of Newtonian gravity.
An experimental test of such a modification of Newtonian gravity would be challenging.

\begin{acknowledgments}
The author gratefully acknowledges the grant from the 
Deutsche Forschungsgemeinschaft (Grant No. La1974/4-1). 
\end{acknowledgments}

\end{document}